\documentclass[a4paper,preprint,12pt]{elsarticle}
\usepackage{amsmath,amsthm,amssymb}
\usepackage{graphicx}% Include figure files
\usepackage{threeparttablex}
\begin{document}
%\linenumbers

\begin{frontmatter}

\title{Comparison of GaN nanowires grown on c-, r- and m-plane sapphire substrates}

%% use optional labels to link authors explicitly to addresses:
\author{Carina B. Maliakkal}
\ead{carina@tifr.res.in}
%\homepage{carina@tifr.res.in}
\author{A. Azizur Rahman}
\author{Nirupam Hatui}
\author{Bhagyashree A. Chalke}
\author{Rudheer D. Bapat}
\author{Arnab Bhattacharya}
%\ead{arnab@tifr.res.in}
%\homepage{arnab@tifr.res.in}

\address{Department of Condensed Matter Physics and Materials Science, Tata Institute of Fundamental Research, Homi Bhabha Road, Mumbai 400005, India.}

\begin{abstract}  Gallium nitride nanowires were grown on c-plane, r-plane and m-plane sapphire substrates in a showerhead metalorganic chemical vapor deposition system using nickel catalyst with trimethylgallium and ammonia as precursors.
% GaN NWs were also obtained on the unpolished backside of c-sapphire used as dummy substrates.
We studied the influence of carrier gas, growth temperature, reactor pressure, reactant flow rates and substrate orientation in order to obtain thin nanowires.
%We observe that the growth rate of the NWs varied inversely with the thickness indicating diffussion-induced growth.??
The nanowires grew along the $<$10$\bar{1}$1$>$ and $<$10$\bar{1}$0$>$ axes depending on the substrate orientation. These nanowires were further characterized using x-ray diffraction, electron microscopy, photoluminescence and Raman spectroscopy.
\end{abstract}

\end{frontmatter}

\section{Introduction}
 Gallium nitride (GaN) nanowires (NWs) have been of interest for nanoscale photonic \cite{Johnson,Duan,Qianled,Qian,Tchernychevaphotonic,Guo,Chenspin}%\cite{Gradecak}
and electronic \cite{Huang,Zhong,Li}%\cite{Huangelectronics,Chenelectronics}
 device applications. The growth of GaN NWs by catalyst mediated \cite{Huang,Zhong,Gottschalch,Zhou,Wang} as well as catalyst free \cite{Guo, Koester, Aschenbrenner,Chae} methods have been reported. Unlike InAs, where $<$111$>$ oriented wires are usually obtained \cite{Hiruma,Jensen,Mandl}, III-nitride NW growth has been observed along different crystal directions and is strongly dependent on growth conditions,  the substrate and its orientation.

In most of the reports of GaN NW growth the axis along the nanowire length (or the growth direction) is along either the polar $<$0001$>$ direction (i.e. c-axis) or the non-polar directions $<$11$\bar{2}$0$>$ (a-axis) and $<$1$\bar{1}$00$>$ (m-axis).
On (0001) c-plane sapphire, Gottschalch $et$ $al.$\cite{Gottschalch} reported the growth of vertical wires  growing along c-axis using gold as catalyst, while Ji $et$ $al.$\cite{JiV} obtained wires growing along the a-axis with nickel catalyst.
On (1$\bar{1}$02) r-plane sapphire using nickel-catalysed  metalorganic chemical vapor deposition (MOCVD), Zhou $et$ $al.$ obtained wires tilted on the substrate and growing along m-axis.\cite{Zhou} On the other hand, Wang $et$ $al.$\cite{Wang} and Qian $et$ $al.$\cite{Qian} reported the growth NWs along the a-axis on r-plane sapphire with nickel catalyst.
The growth of inclined GaN nanorods along the c-axis using m-plane (1$\bar{1}$00) sapphire as substrate without any foreign catalyst was reported recently.\cite{Chae,Tessarek}
%  by Tessarek $et$ $al.$ and Chae $et$ $al.$.
%Apart from the recent report of inclined GaN nanorods grown on m-plane sapphire along the c-axis ($<$0001$>$) without any catalyst \cite{Chae}, there are no reports on the growth of GaN NWs on m-plane sapphire substrates.
There are very few reports on the growth of GaN NWs along semipolar directions.
The growth along the semi-polar $<$10$\bar{1}$1$>$ direction has been reported on c-plane sapphire by Park $et$ $al.$ by reaction of metallic Ga and gaseous NH$_{3}$ by using nickel as the catalyst.\cite{EPark} Peng $et$ $al.$ also reported growth along $<$10$\bar{1}$1$>$ but on graphite substrates, using a mixture of Ga$_{2}$O$_{3}$ and C along with NH$_{3}$ as precursor, by a hot-filament CVD, without any catalyst.\cite{Peng}
%A study comparing GaN NW growth on different substrate orientations will give more insight to the mechanism about their growth, but there are hardly any reported.Sir, better dialogue?
%met graphite, where the growth direction was seen to be temperature dependent.\cite{Peng,EPark}
Tessarek $et$ $al.$ obtained GaN wires via a self-catalyzed method by MOCVD, on different planes of  sapphire, namely c-, r- a- and m-plane. These wires grew invariably along the c-axis with diameters more than 200 nm.\cite{Tessarek}
There are, however, hardly any reports that compare GaN NW growth on different substrate orientations in the same run.
We report the growth of thin, nickel-catalysed, GaN NWs on c-, r- and m-plane sapphire substrates under identical conditions.
We studied the dependence of growth direction on the substrate orientation with other growth conditions kept similar, and characterized the samples using electron microscopy, x-ray diffraction, photoluminescence and Raman spectroscopy.

\section{Growth of nanowires}
GaN NWs were grown using MOCVD with nickel catalyst and trimethylgallium (TMGa) and ammonia (NH$_{3}$) as precursors. The substrates were cleaned, drop-coated with nickel nitrate hexahydrate solution ($\sim$~0.01~M), blow-dried with N$_{2}$ gas  and annealed in hydrogen to form metallic nickel nanoparticles which subsequently served as the catalyst particle.\cite{Wang,Brockner} This method to form the catalyst particles is more convenient than techniques involving evaporation of nickel or gold. The size of the catalyst particle can be controlled by the annealing time and temperature (See Supplementary Information section II for details). Since the exact composition and phase (solid/liquid) of the nickel-gallium alloy that serves as the catalyst during growth is unknown, the NW growth mechanism could be either vapour-liquid-solid (VLS) or vapour-solid-solid (VSS).
%\cite{Zhou}
From a post-growth compositional analysis of the catalyst particle (not presented here), we believe that at the growth temperatures used in our experiments ($\sim840\;^\circ$C) GaN NW growth happens via a VSS process.
We also tried to grow GaN NWs using gold nanoparticles as catalyst. Gold catalysed wires grew slower than their nickel assisted counterparts under similar growth conditions, as reported by Zhou $et$ $al.$.\cite{Zhou}

\begin{figure*}
\includegraphics[width=1\textwidth]{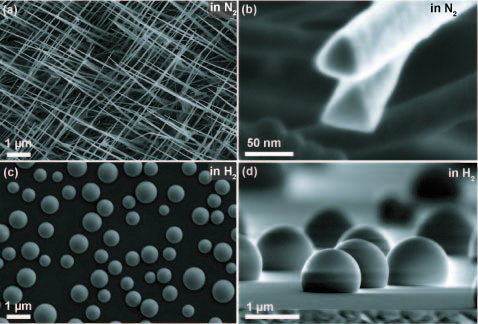}
\caption{ \label{fig:growth} \textbf{Growth of GaN NWs}: (a) GaN NWs obtained by growing in N$_{2}$ environment ($840\;^\circ$C and 150~torr with $\sim$45~$\mu$mol/min NH$_{3}$ flow and 9.7~$\mu$mol/min TMGa flow, on r-plane sapphire). (b) Image showing that the cross section of the NWs grown in N$_{2}$ are triangular. (c) Gallium droplets obtained at  similar growth conditions but under H$_{2}$ ambient. (d) Cross sectional image of the sample shown in (c).}
\end{figure*}

We varied the temperature, reactor pressure and rate of flow of precursors in order to obtain thin and non-tapering NWs. The growth of NWs was sensitive to the reactor conditions. We first discuss the effect of carrier gas on growth. After annealing the nickel nitrate hexahydrate coated sapphire substrates in a hydrogen environment to produce nickel nanoparticles, we could grow GaN NWs in a pure nitrogen ambient. The scanning electron micrograph (SEM) of the NWs is shown in Figure \ref{fig:growth}(a). The wires typically had a triangular cross section (Figure \ref{fig:growth}(b)). However, under similar reactor conditions in an H$_{2}$ ambient or in a mixture of equal volumes of N$_{2}$ and H$_{2}$, we did not obtain any wires. Instead, we obtained just hemispherical gallium droplets (Figure \ref{fig:growth}(c),(d)), unlike earlier reports.\cite{Li}

\begin{figure*}[!ht]
\centerline {\includegraphics[width=1\textwidth]{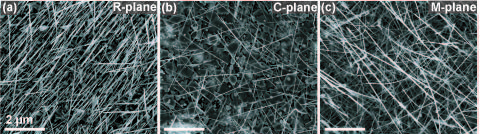}}
\caption{ \label{fig:orientation}
Nanowires on different planes of sapphire: SEM images of nickel-catalysed non-tapering GaN NWs obtained in the same growth run on (a) r-plane (b) c-plane and (c) m-plane sapphire substrates.
(Growth conditions: $840\;^\circ$C and 150~torr with $\sim$45~$\mu$mol/min NH$_{3}$ flow and 7.8~$\mu$mol/min TMGa flow, for 40 minutes)}
%The scale bar in all the images corresponds to 2~$\mu$m.}
\end{figure*}

\begin{figure*}[!ht]
\centerline {\includegraphics[width=0.5\textwidth]{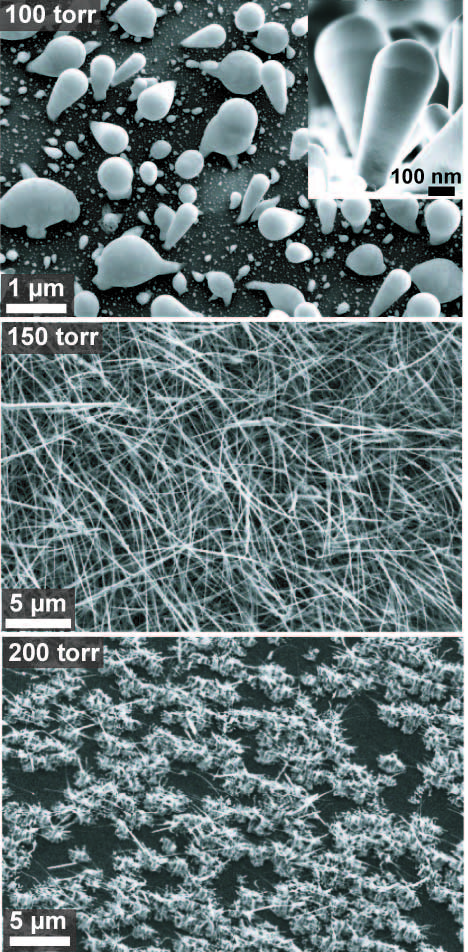}}
\caption{ \label{fig:pressure}
Effect of pressure: SEM images of sample grown at different ambient pressure on r-plane sapphire substrates at $840\;^\circ$C with $\sim$45~$\mu$mol/min NH$_{3}$ flow and 9.7~$\mu$mol/min TMGa flow.}
%The scale bar in all the images corresponds to 5~$\mu$m.
\end{figure*}

NWs were grown on different orientations of sapphire namely c-plane, r-plane and m-plane. The NWs obtained from the same growth run at $840\;^\circ$C and 150~torr with $\sim$45~$\mu$mol/min ($\sim$1~sccm) NH$_{3}$ flow and 7.8~$\mu$mol/min (2~sccm) TMGa flow on these substrates are shown in Figure \ref{fig:orientation}. On these substrates we obtained NWs as thin as 20~nm diameter with a triangular cross section.  The growth temperature was varied between $820\;^\circ$C and $1020\;^\circ$C. Experiments carried out at different reactor pressures between 50~torr and 200~torr (Figure \ref{fig:pressure}) showed that long thin NWs were obtained at 150~torr. At a relatively low pressure of 100~torr icecream-cone shaped structures were formed. A lot of wires grown at 175 torr had kinks or had a zig-zag morphology. At 200~torr there were very little NW growth, and those few were highly tapered and short.
 %On decreasing ... On increasing .....
 Very low flow of precursors (7.8-9.7~$\mu$mol/min of TMGa and $\sim$45~$\mu$mol/min of NH$_{3}$) and small V/III ratio ($\sim5$) was used to facilitate anisotropic growth. At large V/III ratio the growth is slower, while at low V/III ratio the NWs obtained are more tapered. (For SEM images  refer Supporting Information Section I.)
 We found that a pressure of 150 torr in N$_{2}$ environment, $\sim~840\;^\circ$C and a V/III ratio of $\sim5$ was optimum for obtaining thin non-tapering wires.

%We use a susceptor with 2" pockets to place the substrate wafers. For NW growth we were using substrate pieces of about 1X1" in size. In order to avoid any deposition on these pockets we used kept c-sapphire substrates with their backside facing up in these pockets, and the growth substrates on top of them. We found good NW coverage on these wafers on the part close to the samples intensionally coated with catalyst. We suspect that the Ni catalyst on the substrates spilled over and lead to NW growth on them.

A comparison of the NWs obtained on different planes which are shown in Figure \ref{fig:orientation} is given in Table 1. All these samples compared are from the same growth run (at $840\;^\circ$C and 150~torr with $\sim$45~$\mu$mol/min NH$_{3}$ flow and 7.8~$\mu$mol/min TMGa flow). The length of the NWs were determined by imaging cross sectional samples.
The NWs obtained on c-plane sapphire are narrower than the NWs on r-plane. On m-plane sapphire we see two distinct set of NWs -- one which is narrow (thickness $\sim$~30~nm) and one which is thicker (thickness $\sim$~90~nm). Since the NWs on m-plane sapphire are not erect, it is difficult to measure its exact length, but they are longer than NWs grown on c- and r-plane.

\begin{table}[!ht]
\label{tab:compare}
\vspace{0.2cm}
\centering
\begin{tabular}{lcccc}
Sapphire          & Nanowire                                & Average (mode)    & Average (mode)  & Low temp. \\
substrate          & growth                                &of length             &of diameter      & PL FWHM \\
orientation        & direction                             & ($\mu$m)             &  (nm)          &  (meV)  \\
\hline \\[-1ex]
r-plane            & $<$10$\bar{1}$0$>$                       & 2.5                & 40           &   390   \\
c-plane            & $<$10$\bar{1}$0$>$                    & $\sim$3               & 25           &   220   \\
m-plane            & $<$10$\bar{1}$1$>$, $<$10$\bar{1}$0$>$  & $\sim$13            & 30, 90        &   350   \\

\end{tabular}
\vspace{0.5cm}
\caption{Comparison of NWs grown on different orientations of sapphire under similar conditions. The average length and diameter of the NWs given in the table are from the same run. (Growth conditions: $840\;^\circ$C and 150~torr with $\sim$45~$\mu$mol/min NH$_{3}$ flow and 7.8~$\mu$mol/min TMGa flow. The SEM images of these wires has been shown in Figure \ref{fig:orientation})}
\end{table}

\section{Structural characterization of nanowires}
The crystal structure and the crystallinity of the NWs were analyzed using x-ray diffraction (XRD) and transmission electron microscopy (TEM).
\subsection{XRD analysis}

\begin{figure*}[!ht]
\centerline {\includegraphics[width=0.5\textwidth]{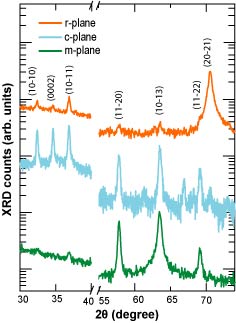}}
\caption{ \label{fig:xrd}
X-ray diffraction pattern with the planes indexed showing that the GaN wires have crystallized in the wurtzite phase.(XRD plots are displaced for clarity.)}
\end{figure*}

The grazing incidence XRD pattern the GaN nanowires grown on c-plane, r-plane and m-plane sapphire substrates is shown in Figure \ref{fig:xrd}. The peaks are quite sharp indicating good crystal quality. The peak positions have been indexed to the wurtzite crystal structure of GaN.
The lattice constants obtained by least square fitting from the peak positions were a~=~3.188~{\r{A}} and c~=~5.179~{\r{A}}which agrees with values reported in literature.\cite{Leszcynski, Vurgaftman}

% The strong diffraction peaks relative to the background signals suggested that the resulting GaN nanowires had a high purity of the GaN (plagiarism sentence) wurtzite phase.

\subsection{TEM analysis}

\begin{figure*}[!ht]
\includegraphics[width=1.0\textwidth]{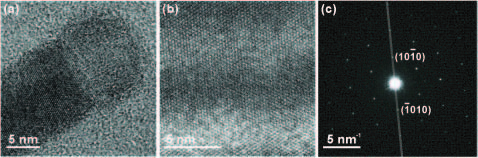}
\caption{ \label{fig:tem} \textbf{TEM of NWs}: (a) TEM image showing a catalyst particle at the tip of the NW indicating that the growth mechanism is VLS or VSS. (b) High resolution TEM image of a NW grown on c-plane sapphire. (c) Electron diffraction pattern of the same wire as in (b)viewed along the [0001] zone axis, where the line indicates axis of wire growth. We hence infer that the wires grown on c-plane sapphire grow along $<$10$\bar{1}$0$>$ i.e. m-axis.}
\end{figure*}

\begin{figure*}[!ht]
\centerline {\includegraphics[width=0.5\textwidth]{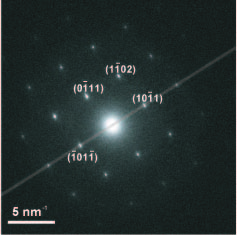}}
\caption{ \label{fig:temms}
 %\textbf{TEM of NWs grown on m-plane sapphire}:
Electron diffraction pattern of a NW grown on m-plane sapphire when viewed along the [$\bar{1}$101] zone axis. The line indicates axis of NW growth.  This particular wire is oriented along the $<$10$\bar{1}$1$>$ direction.}
\end{figure*}

The crystallinity and growth direction of the wires grown on c-, m- and r-plane sapphire were determined using TEM. Since the NW thickness is less than 100~nm they are electron transparent and do not require further thinning down for TEM studies. The sample on which wires were grown was ultrasonicated in methanol to make a suspension. This suspension was slowly dropped multiple times onto a TEM grid and the solvent allowed to evaporate. The presence of the catalyst nanoparticle seen at the end of the wire in the TEM image (Figure \ref{fig:tem} (a)) confirms that the growth was catalytic. From energy dispersive X-ray spectroscopy the presence of Ga in the catalyst along with Ni was confirmed. The lattice planes seen in the high resolution TEM images (Figure \ref{fig:tem}(b)) and the well-defined electron diffraction pattern obtained (Figure \ref{fig:tem}(c)) confirm the single crystal nature of these wires. The diffraction pattern was indexed to find the lattice planes that gave rise to the diffraction spots. By comparing the diffraction pattern and the low magnification TEM image of the wire, the growth axis was determined. An indexed electron diffraction pattern obtained with a TEM is shown in Figure \ref{fig:temms} showing that the growth direction is $<$10$\bar{1}$1$>$ for this wire. The wires grew along the m-axis ($<$10$\bar{1}$0$>$) on both c-plane (Figure \ref{fig:tem}(c)) and r-plane sapphire substrates. On m-plane sapphire most of the NWs grew along the $<$10$\bar{1}$1$>$ (Figure \ref{fig:temms}) and $<$10$\bar{1}$0$>$ directions (Table 1).

\section{Optical characterization}

The optical properties of the GaN NWs were studied using photoluminescence (PL) spectroscopy from ensembles of NWs. The sample was excited with a 266~nm frequency-quadrupled Nd:YAG laser. The luminescence from the sample was dispersed through a 0.55~m monochromator and detected with a thermoelectrically cooled Si-CCD.
\begin{figure}[!ht]
\centerline {\includegraphics*[width=0.5\textwidth]{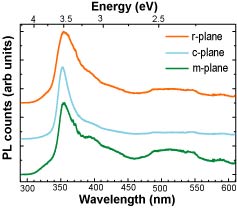}}
\caption{ \label{fig:PL} \textbf{PL from GaN NWs}: PL spectra of ensembles of NWs at 10~K that were grown on c-plane, r-plane and m-plane sapphire substrates and mechanically transferred to an Si/SiO$_{2}$ substrate. (Spectra are displaced for clarity.)}
\end{figure}
The PL spectra of GaN NWs at 10~K grown on c-plane, r-plane and m-plane sapphire are shown in Figure \ref{fig:PL}. The spectra are normalized such that the near-band-edge emission peak is of the same intensity. PL spectra from these NWs peak around 3.50~eV (354~nm), which corresponds to the near-band-edge emission of GaN.\cite{JiV,Monemar,Robins} The FWHM of the PL peaks are $\sim$~22 nm (220~meV), $\sim$~42~nm (390~meV) and $\sim$~37~nm (350~meV) for NWs grown on c-plane, r-plane and m-plane sapphire respectively. These FWHM values are comparable to those reported in literature.\cite{Lyu,chin} The width of these spectra could be due to the variation in thickness and properties of the different wires of the ensemble,\cite{JiV} or due to defect-related luminescence.\cite{Reshchikov}
 %refer Monemar for bulk 3.053 eV XX 0.01 eV FWHM
 %Ji 3.471 eV NBE in NW Ni MOCVD
%Robins NW cat.free 3.48 eV
% width cite Lyu, Chin
  %The dispersion in sizes of the NWs might be giving rise to the relatively broad PL spectrum.
  % The PL peaks from these NWs peak around  3.50 eV (354 nm) which are slightly blueshifted compared to bulk GaN emission with a peak (3.47 eV or 357 nm) at the same temperature (10 K).
   %This blueshift could be arising due to several reasons like strain in the wire, quantum confinement effects, \cite{Ha} surface traps \cite{Chin} and dependence of bandstructure on polarization fields \cite{Kuykendall}. The luminescence from very thin (diameter comparable to Bohr exciton radius, $\sim$ 11 nm for GaN) NWs are expected to be blueshifted and stronger that from thicker wires \cite{Ha}.
  % In the sample for which PL was recorded the sides of the wires were ranging from $\sim$ 20 to 60 nm  XXXX.
%This peak vanishes as the temperature of the sample is increased.
 %they had sharp peak & can at high excitation only
An additional peak is observed at $\sim$ 3.1~eV (400~nm) which might be due to strongly localized excitons.\cite{Reshchikov}. Yellow luminescence from these samples is relatively small indicating good crystal quality.
 %The room temperature PL of wires grown on m-sapphire as well as c-sapphire peaks around  355 nm as seen in  Fig. \ref{fig:CLPLRaman}.a. The peaks are slightly blueshifted compared to bulk GaN emission with a peak at 363 nm at the same temperature. This blueshift could be arising due to several reasons like strain in the wire, quantum confinement effects, \cite{Ha} surface traps \cite{Chin} and dependence of bandstructure on polarization fields \cite{Kuykendall}. The luminescence from very thin (diameter comparable to Bohr exciton radius, $\sim$ 11 nm for GaN) NWs are expected to be blueshifted and stronger that from thicker wires \cite{Ha}. In the sample for which PL was recorded the sides of the wires were ranging from $\sim$ 20 to 60 nm. The large dispersion in NW size also results in a broad PL spectrum with FWHM values of $\sim$ 43 nm and $\sim$ 35 nm for NWs on c-plane and m-plane sapphire respectively, similar to earlier reports \cite{Chin,Lyu}. Photoluminescence spectra from bare sapphire substrates shows a peak at $\sim$ 380 nm, which also contribute to the width and asymmetry of the PL spectra of the sample with wires.

%\section{CL}

\section{Raman spectroscopy of NWs}
%"Phonon modes of GaN have received considerable attention because information on them is important in considering the~electron transport, the non-radiative electron relaxation process, and so on"

Raman spectroscopy was used to study the phonon modes of GaN NWs. A Witec alpha 300R confocal Raman microscope was used with a 532~nm frequency doubled Nd:YAG laser for excitation.
%Raman spectra were obtained using a Witec alpha 300R confocal Raman microscope using a 532~nm frequency doubled Nd:YAG laser for excitation.
\begin{figure}[!h]
\centerline {\includegraphics[width=0.5\textwidth]{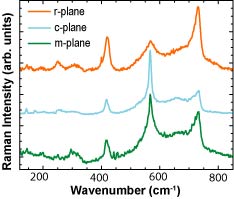}}
\caption{ \label{fig:Raman} \textbf{Raman spectra of GaN NWs}: Raman spectrum of a individual NWs grown on r-plane, c-plane and m-plane sapphire substrates. (Spectra are displaced for clarity.)}
% and transferred to CaF$_{2}$ substrate.}
\end{figure}
Figure \ref{fig:Raman} shows the Raman spectra for single GaN NWs, that were transferred to an aluminium foil which serves as the substrate. Al foil was chosen as substrate since it provides much lesser background signal in the region of interest than conventional crystalline substrates like Si, GaAs, sapphire, etc and other amorphous substrates.
The intensity of the Raman peaks depends on the direction of both the incident and scattered polarization vectors relative to the crystal orientation. In order to observe as many phonons as possible from NWs from all samples the incident electric field was set at an angle of $45^\circ$ with the NW axis, and the scattered light of all polarizations were detected simultaneously. (The spectra shown is after background correction.)
Difference in relative intensity may due to difference in growth direction of NW and/or because the incident light might be falling in different direction relative to the crystallographic directions.

At around 144, 567 and 734~cm$^{-1}$ the E$_{2}$ (low), E$_{2}$ (High) and E$_{1}$ (LO) are seen.\cite{azuhata_polarized_1995,Davydov,bungaro_textitab_2000}
%531 hidden A1 (TO) ?
These peaks are allowed by group theory in a infinitely large sample with wurtzite crystal structure belonging to the space group $C^4_{6v}$ (P6$_{3}$mc), and has been observed experimentally in first order Raman spectra.
But, the other peaks observed in the spectrum shown (Figure \ref{fig:Raman}) arise due to the finite dimensions of the NWs.
The acoustic phonons, and also the combination of both acoustic and low-lying optical branches at the M symmetry point in the Brillouin zone gives rise to the spectral peak around 256~cm$^{-1}$.\cite{chen_catalytic_2001}
The peak at $\sim$ 308~cm$^{-1}$ arises from the overtone process of acoustic phonons and the energy position matches the flat phonon branch at the H-point in the Brillouin zone.
The peak observed in the spectra at $\sim$ 418~cm$^{-1}$ also is an acoustic overtone corresponding to the M-point.\cite{Ha,Siegle,chen_catalytic_2001,Davydov}
%"415-'the combination of optical and acoustic branches at the M point" davydov'.
% chen also calls 250 'zone boundary phonon'
%The broad peak at 666~cm$^{-1}$ could be either a surface mode (??check in bulk use some pressed sample, with different ambient)or might be arising from defect levels \cite{livneh_polarized_2006}) (check with 633 laser??)
%The broad peak at 666~cm$^{-1}$ arise from vacancy-related defects.\cite{livneh_polarized_2006,chen_catalytic_2001}
The Raman peaks are broader compared to those reported in bulk samples due to the finite size effects in accordance with the phonon confinement model, where the spatial confinement of phonons in nanocrystals leads to an uncertainty in wave vector and hence a broadening of the Raman spectral lines.\cite{richter_one_1981,kanata_raman_1987,chen_catalytic_2001}

\section {Conclusions}
In conclusion, we have grown GaN NWs on c-plane, m-plane, and r-plane sapphire substrates using nickel catalyst and TMGa and NH$_{3}$ as precursors by MOCVD.
% We also obtained NWs on the unpolished backside of c-sapphire substrates.
A pressure of 150~torr in N$_{2}$ environment and $\sim~840~^\circ$C, with low precursor flow rates and V-III ratio, was optimum for growing thin, non-tapering GaN NWs. The wires had a triangular cross-section and grew along the $<$10$\bar{1}$1$>$ and $<$10$\bar{1}$0$>$ directions. Low temperature PL shows near-band-edge emission from the NWs and small yellow luminescence indicates good crystalline quality. Raman spectra reveals the good crystal quality of these NWs and effects due to finite size.
% We see that the growth rate of the NWs varies inversely with the thickness indicating diffussion-induced growth??.

\section {Acknowledgments}
The authors are thankful to Mandar Deshmukh, Priti Gupta, Sandip Ghosh, Ritam Sinha and S.C. Purandare at TIFR, India and Subramaniyam Nagarajan at Aalto University, Finland for support in materials characterization and useful discussions.
This work at TIFR was supported through internal grants 12P0168 and 12P0169. C.B.M. acknowledges travel support from the Department of Science and Technology, India, through the collaborative project INT/Finland/P-10.

\section*{References}

%\bibliographystyle{unsrt}
%\bibliography{GaNNWpaper}
%\end{document}

\end{document}